# Existence of long-lived isomeric states in naturally-occurring neutron-deficient Th isotopes

Amnon Marinov[1], Ilia Rodushkin[2], Yoav Kashiv[1], Ludwik Halicz[3], Irena Segal[3], Arthur Pape[4], Robert V. Gentry[5], Harold W. Miller[6], Dietmar Kolb[7] & Reinhard Brandt[8]

[1] *Racah Institute of Physics, The Hebrew University of Jerusalem, Jerusalem 91904, Israel*
[2] *Analytica AB, Aurorum 10, S-977 75 Luleå, Sweden*
[3] *Geological Survey of Israel, 30 Malkhei Israel St., Jerusalem 95501, Israel*
[4] *IPHC-UMR7500, IN2P3-CNRS/ULP, BP28, F-67037 Strasbourg cedex 2, France*
[5] *Earth Science Associates, P.O.Box 12067, Knoxville, TN 37912-0067, USA*
[6] *P.O.Box 1092, Boulder, CO 80306-1092, USA*
[7] *Department of Physics, University GH Kassel, 34109 Kassel, Germany*
[8] *Kernchemie, Philipps University, 35041 Marburg, Germany*

**Evidence for the existence of long-lived neutron-deficient isotopes has been found in a study of naturally-occurring Th using inductively coupled plasma-sector field mass spectrometry. They are interpreted as belonging to the recently discovered class of long-lived high spin super- and hyperdeformed isomers.**

In recent years, long-lived high spin super- and hyperdeformed isomeric states with unusual radioactive decay properties have been discovered in heavy and very heavy nuclei[1-4]. This discovery motivated us to perform a search for naturally-occurring long-lived isomeric states. (Up to now there is only one such isomeric state known, namely the 75.3 keV excited state in $^{180}$Ta with a half-life of $>1.2 \times 10^{15}$ y (ref.[5])). Madagascan monazite and commercially available Th and U standard solutions were studied using different mass spectrometers, including an accelerator mass spectrometer. In the present paper we present the results obtained with pure Th standard solutions and also in one case with a monazite digest solution, using inductively coupled plasma-sector field mass spectrometer (ICP-SFMS).

The instrument was an Element2 (Thermo-Electron, Bremen, Germany). The predefined medium resolution mode of m/Δm = 4000 (10% valley definition) was used throughout the experiments so as to separate atomic ions from interfering molecular ions with the same mass number. The sensitivity-enhanced setup of the instrument was similar to that described in Ref.[6] where a capacitive decoupling system and high-performance "X" skimmer were used, providing sensitivity for $^{232}$Th in the medium resolution mode of up to $2 \times 10^8$ counts $s^{-1} mg^{-1} l^{-1}$. The sample uptake rate was approximately 60-80 μl $min^{-1}$ (controlled by a PFA Teflon micronebulizer operated in the self-aspiration mode). Methane gas was added to the plasma to decrease the formation of interfering molecules[7]. Oxide and hydride formation (monitored as $UO^+/U^+$ and $UH^+/U^+$ intensity ratios) were approximately 0.04 and $1 \times 10^{-5}$, respectively. Accurate mass calibration was performed daily using the $^{209}Bi^+$, $^{232}Th^+$, $^{235}U^+$, $^{238}U^+$ and $^{238}U^{16}O^+$ peaks. Two 1000 mg $l^{-1}$ Th stock solutions "A" and "B" from two manufacturers, Inorganic Venture and Customer Grade, were obtained from LGC Promochem AB (Borås, Sweden). Complete elemental screening was performed on both solutions to assess the impurity concentration levels[*]. The following concentrations (expressed as ppm (parts per million) of the Th concentration) of certain elements that can potentially give rise to spectrally interfering species were found:
A: U 80, Bi 0.01, Pb 0.2, Hg 0.03, Au 0.0004, Hf 0.02, Ce 0.7, Dy 7, B 3, Be 0.03.
B: U 2200, Bi 14, Pb 300, Hg 0.2, Au 0.006, Hf 0.02, Ce 8, Dy 0.7, B 3, Be 0.4.

---

[*] The concentration level of various impurities might be different in different batches from the same manufacturer.

The solutions were analyzed during three separate sessions: May 25, October 5, and November 6, 2005. A range of about 0.2 mass unit was scanned in each measured spectrum. This range was divided into approximately 60 channels. During the first session, masses from 210 to 269 were analyzed with an integration time per channel of 0.6 sec. During the second and third sessions, selected mass regions (where some indication of unidentified signals had been detected) were measured using an integration time per channel of 3 and 12 sec, respectively. Instrumental sensitivity varied significantly among runs as a result of matrix effects caused by the introduction of highly concentrated solutions into the ICP source. During the first session, the monazite digest (1000 mg monazite $l^{-1}$) and Th solution A (diluted to 20 mg Th $l^{-1}$) were scanned once. (The Th content in the monazite was approximately 2%. The contents of typical rare earth elements in it, like Ce, Dy and Er, were about 5%, 0.05% and 0.02%, respectively.) During the second session, 20 mg $l^{-1}$ of Th A and B solutions spiked with 2 µg $l^{-1}$ Bi were studied and each solution was measured three times. The Th concentration was increased to 50 mg $l^{-1}$ during the last session, and both solutions were measured twice. Altogether eleven spectra were taken for each mass number studied with the Th solutions. Replicate analyses of blank solution (0.14 M $HNO_3$) were performed. Very few events appeared in the blank measurements, and no event appeared in the mass regions that correspond to atomic masses.

Figure 1 shows summed spectra of six measurements performed during the second session in the regions of the $^{209}Bi^+$, $^{230}Th^+$ and $^{238}U^{16}O^+$ peaks. The results show an accuracy of the mass calibration of approximately 0.002 mass units, using the $^{209}Bi^+$ peak for on-line adjustment of the mass calibration (so-called "lock-mass" feature of the ICP-SFMS). However, shifts up to about 0.025 mass units were sometimes seen in the measurements. (These shifts were sometimes corrected by using interference peaks present in the spectra. The maximum correction that was applied in the data presented in this paper was 0.013 amu.) The full width at half maximum (FWHM) of the peaks is about 0.030 mass units.

Figure 2 (left) presents the results of three measurements on mass 210, one measurement on the Th solution A from the first session and one measurement each on solutions A and B from the third session. In addition to the identified interference peaks observed at $M_{exp.}$(c.m.) = 209.934 amu in Fig. 1b and at $M_{exp.}$(c.m.) = 209.976 amu in Fig.1c, each of these measurements also shows one event (indicated with an arrow) at an average mass of 210.021. (In Fig. 2c, the experimental value was corrected according to the shift in the interference peak identified as $^{209}Bi^1H^+$.)

Figure 2 (right) displays the results of two measurements on mass 211 obtained on solution B during the third session, where two events were detected in the first measurement and four events in the second, at a weighted average mass of 211.024.

Figure 3 shows five results on mass 213 and one result on mass 214. In Fig. 3a, two events were detected at mass 213.021 during the first session. During the second session (Figs. 3b and 3c) four counts (solution A) and one count (solution B), respectively, are seen at a weighted average mass of 213.006. Figures 3d and 3e show the results of two measurements on solution A obtained during the third session where one event was detected in each measurement at an average mass of 213.019. (In these cases the measured masses were corrected according to the shifts found in the identified interference peaks.)

Figure 3f present the results obtained during the third session at mass 214 on solution A. In addition to the two events related to an identified molecular ion at lower mass (213.953) in the figure, one count was detected at 214.022.

The results for mass 217 are given in Fig. 4. Figure 4a shows a spectrum from solution A obtained during the first session. In addition to a few background counts, six events were detected at a center of mass (c.m.) of 217.027. Figure 4b sums the six measurements from the second session shown in Fig. 5. A cluster of seven counts is seen at a c.m. value of 217.011 (We assume here that the event at mass 217.001 amu could be a background event.) Some background counts are seen in this figure both in the lower mass region (probably due to molecular interferences) and also at higher mass around 217.060. It is seen in Fig. 5 that the events at mass 217.060 appeared in one of the six spectra only (Fig. 5a). We therefore assume that this may be due to accidental ion scattering. In contrast to the other spectra obtained during the present study, where background counts other than those originating from identified interferences were rare, here the background is substantial. However, the cluster at mass 217.010 is rather striking. The probability that 7 out of a total number of 26 randomly distributed counts will occur accidentally in a predefined small region of 3 channels out of a total region of 60 channels is small, about $2 \times 10^{-4}$. This value ($P_{acc.}$) is calculated from

$$P_{acc.} = \binom{N}{n} \left(\frac{r}{R}\right)^n \left(1 - \frac{r}{R}\right)^{N-n}$$

where $N$ is the total number of events in the whole measuring region consisting of $R$ channels, and $n$ is the number of events in a subset of $r$ sequential channels.

In Fig. 4c, two counts are seen at a c.m. of 217.022. However, these two events are not clearly distinguishable from background.

Figure 6 presents the results obtained on mass 218. Figures 6a and 6b show spectra obtained in the first session with pure Th and with monazite solutions. One event is seen in the first plot at mass 218.012 and one in the second plot at mass 218.020. (The correction in the latter case is based on the observed shift in the mass position of the interfering $^{162}Dy^{40}Ar^{16}O^+$ molecule (see Fig. 6c)**. Two counts were obtained at an average mass of 218.013 with solution B during the second session as seen in Fig. 6d. Five counts that repeated themselves (Figs. 6e and 6f) were seen with solution A during the third session at average masses of 218.029 and 218.021, respectively. The measured masses were corrected in these cases for the shifts observed in the $^{208}Pb^+$, $^{230}Th^+$ and $^{238}U^{16}O^+$ peaks, which were measured in the same sequences.

The results are summarized in Table 1. We note that the events assigned to a particular mass as obtained in the different experiments always occurred within the measured FWHM of 0.030 amu. We have been unable to match the observed signals to any known molecular ions. Usually, as seen in some of the plots, the masses of the interfering molecules are located below the masses of the new peaks reported here. One possibility is that the four counts at mass 213.006 (Fig. 3b) may be due to $BiH_4^+$ that has a molecular mass of 213.012. However, the intensity of $BiH^+$ was measured to be about $1.1 \times 10^{-5}$ of the $Bi^+$ signal. For the integration time used to obtain the results of Fig. 3b and Fig. 1a, this would correspond to approximately 6 counts. Even if it is assumed that the probability for adding a second, third and fourth $H^+$ ions is somewhat larger than $1 \times 10^{-5}$ in each step, it is clear that the intensity of $BiH_4^+$ (on the order of $6 \times 10^{-15}$ counts) should not be detectable at the Bi concentration present in the solution. Another possibility that should be considered is the potential presence of hydrocarbon-based molecular ions from pump oils. Their typical masses are, however, well separated from and higher than the masses of the events seen. For instance, the mass of $CH_3(CH_2)_{14}^+$ is 211.243, to be compared with an average mass of 211.024 which was observed in this case.

On the other hand, it is evident from the data presented in Table 1 that the average masses of the measured peaks fit the known ground state (g.s.) masses[8] of several Th isotopes to within 0.01 mass unit.

The intensities of the observed events do not always follow the dwell time of the various measurements and some of the spectra (out of the eleven measured for each mass number) even did not yield events for certain Th masses (see Table 1, Column 3). This could be due to the extremely low counting rates resulting in poor statistics, as well as primary differences in trace isotope contents of the different solutions. It is estimated that the concentration of these minor Th isotopes amounts to $(1-10) \times 10^{-11}$ of $^{232}$Th (or about $(2-20) \times 10^{-16}$ of the solutions).

---

** Another small interfering peak at $M_{exp}$ (c.m.) = 217.912 amu that fits the mass of $^{170}Er^{16}O_3$ (M = 217.920 amu[8]) is seen in this figure.

The half-lives of the corresponding known Th isotopes in their g.s. are very short, on the order of tens of msec to less than 1 μsec[5]. This prompts us to conclude that the observed Th events are due to previously unknown long-lived isomeric states. (The accuracy of the present experiment is not enough to determine the excitation energies of the isomeric states.) If their concentration in Earth was originally similar to that of $^{232}$Th, then the lower limit on their half-lives would be about $1 \times 10^8$ y.

The character of the observed isomeric states is not clear. They cannot represent the high spin isomers that occur near closed shells, as the observed isotopes are far from closed shells. Nor are they related to the fission isomers found in actinide nuclei, since their lifetimes are in the region of msec to nsec. In principle they could be high spin K-type isomers[9] where hundreds of them are known. However, the lifetimes of almost all of them (except of the isomeric state in $^{180}$Ta mentioned above) are very short compared to $1 \times 10^8$ y. It is reasonable to assume that the isomeric states seen in the present experiment are high spin (possibly K-type) super- and hyperdeformed isomeric states like those mentioned above[1-4], where the high spin, the barriers between the various minima of the potential-energy surfaces and the unusual radioactive decay properties contribute to the long lifetime.

The nucleosynthesis process(es) that leads to the production of the observed isomeric states is not known. High spin states in general and such states in the super- and hyperdeformed minima in particular are preferentially produced by heavy ion reactions[4,10]. The observed isomeric states suggest that heavy ion reactions are involved in their production.

In summary, evidence for the existence of long-lived isomeric states with $t_{1/2} \geq 1 \times 10^8$ y has been found for the neutron-deficient isotopes $^{210}$Th, $^{211}$Th, $^{213}$Th, $^{217}$Th and $^{218}$Th, and possibly for $^{214}$Th as well. They are interpreted as high spin super- and hyperdeformed isomeric states. The discovery of such nuclides in natural (non-irradiated) Th lends independent support to the recently proposed coherent description of previously unexplained radioactivities[11] by such isomeric states and to a new outlook on the possible existence of superheavy elements in nature[10].

**Acknowledgements**

We appreciate valuable discussions and the help of M. Paul, F. Oberli, N. Zeldes, S. Gelberg, O. Marinov and H. Feldstein. D. K. acknowledges financial support from the DFG.

**Competing interests statement**

The authors declare that they have no competing financial interests.

Correspondence and requests for materials should be addressed to A. Marinov (email: *marinov@vms.huji.ac.il*).

**Table 1 Summary of mass measurements and comparison to the known masses of the various Th isotopes.**

| Mass Number | No. of Events | No. of Measurements[1] | Fig. No. | Total Dwell Time (sec) | $M_{exp.}$[2] (Average) | $M_{g.s.}$ of Th Isotope[3] |
|---|---|---|---|---|---|---|
| 210 | 3 | 3 | 2 | 24.6 | 210.021 | 210.015 |
| 211 | 6 | 2 | 2 | 24.0 | 211.024 | 211.015 |
| 213 | 9 | 5 | 3 | 30.6 | 213.012 | 213.013 |
| 214 | 1 | 1 | 3 | 12.0 | 214.022 | 214.012 |
| 217 | 13(15)[4] | 6(7)[4] | 4 | 30.6 | 217.018 | 217.013 |
| 218 | 13(14)[5] | 4(5)[5] | 6 | 28.2 | 218.021 | 218.013 |

[1] Eleven measurements were performed for each mass number with the Th solutions and one with the monazite solution (see text).
[2] The errors in mass position are estimated to be ±0.015 amu.
[3] Audi, G., et al., Nucl. Phys. A729, 337-676 (2003). (Ref. 8).
[4] The number in parentheses includes the data of Fig. 4c.
[5] The number in parentheses includes the monazite data (Fig. 6b).

# Figure Captions

**Figure 1** Results of mass measurements on $^{209}$Bi, $^{230}$Th and $^{238}$U$^{16}$O obtained in the second session. Sum of six measurements, three with A solution and three with B solution are displayed. $M_{exp.}$ - observed mass position; c.m. – center of mass of the line; M($^A$Z), M($^{A1}$Z1$^{A2}$Z2) – the known mass of the atom or the molecule taken from Ref.[8] (see text).

**Figure 2** Results of measurements for mass regions 210 and 211 obtained from several sessions. Abbreviations: Th_210, Th_211 – Th solution, mass 210, 211, respectively. A – Th solution A; B – Th solution B; 1,2,… – measurement 1,2,… during the same session on the same solution; $M_{exp.}$ – observed mass position; c.m. – center of mass of the line; corr. – value corrected according to the mass shift found for the molecular interference present; M($^A$Z), M($^{A1}$Z1$^{A2}$Z2$_n$) – the known mass of the atom or the molecule taken from Ref.[8] (see text).

**Figure 3** Results of measurements for mass regions 213 and 214 obtained from several sessions (see text and caption to Fig. 2).

**Figure 4** Results of measurements for mass region 217 obtained during several sessions. Figure 4b represents the sum of the six measurements shown in Fig. 5 (see text and caption to Fig. 2).

**Figure 5** Results of measurements for mass region 217 obtained in the second session (see text and caption to Fig. 2).

**Figure 6** Results of measurements for mass region 218 obtained on pure Th solutions and on a monazite solution (see text and caption to Fig. 2). Abbreviation: HS – high scale; Monazite_218 – monazite digest, mass 218.

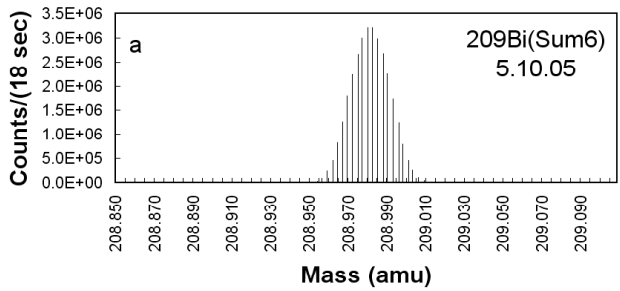

$M_{exp.}(c.m.) = 208.981; \ M(^{209}Bi) = 208.980$

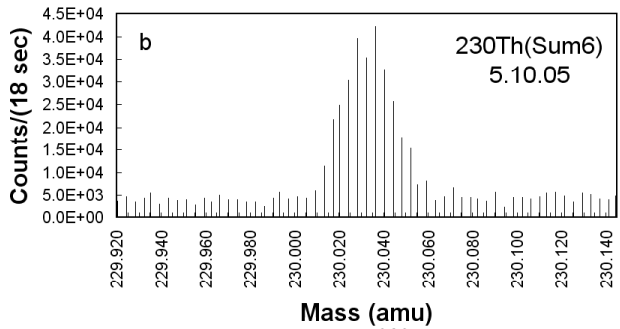

$M_{exp.}(c.m.) = 230.035; \ M(^{230}Th) = 230.033$

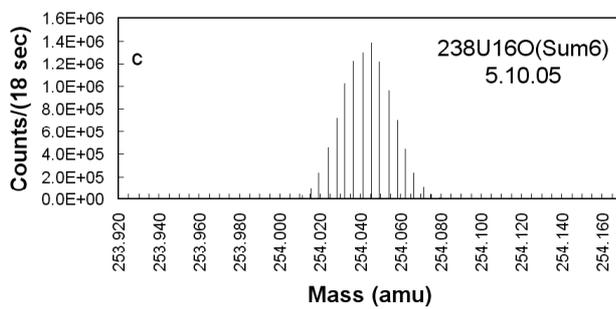

$M_{exp.}(c.m.) = 254.043; \ M(^{238}U^{16}O) = 254.045$

Marinov_fig1

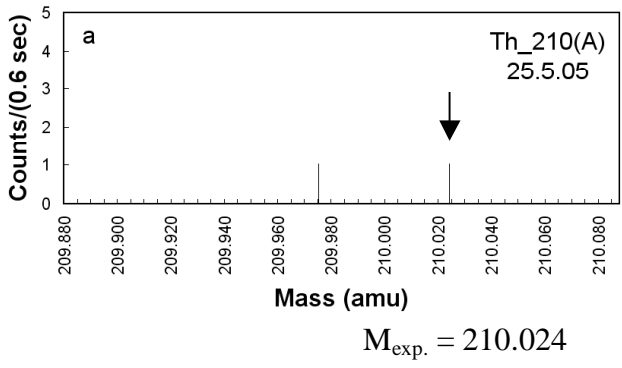

$M_{exp.} = 210.024$

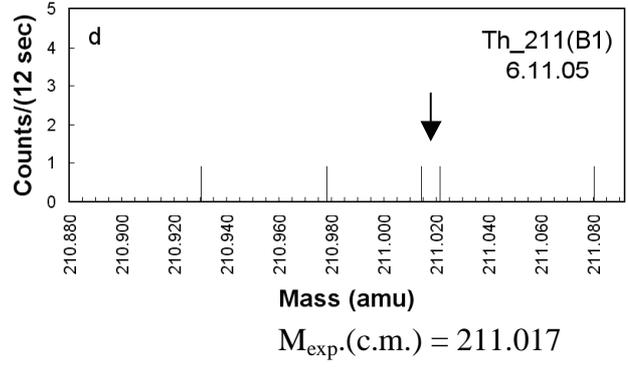

$M_{exp.}(c.m.) = 211.017$

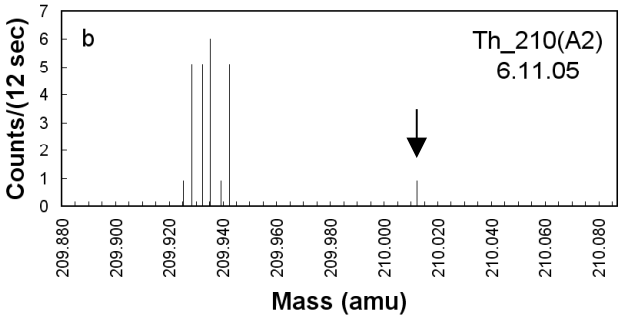

$M_{exp.}(c.m.) = 209.934$   $M_{exp.} = 210.013$
$M(^{178}Hf^{16}O_2) = 209.934$

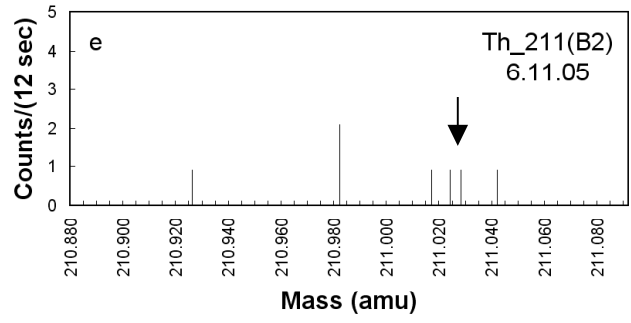

$M_{exp.}(c.m.) = 211.028$

$M(^{211}Th) = 211.015$

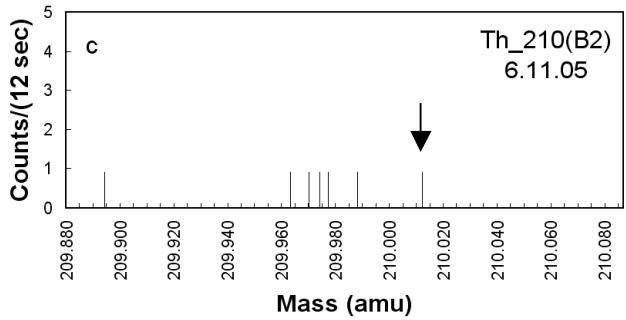

$M_{exp.}(c.m.) = 209.976$   $M_{exp.} = 210.013$
$M(^{209}Bi^1H) = 209.988$   $M_{exp.}(corr.) = 210.025$

$M(^{210}Th) = 210.015$

Marinov_fig2

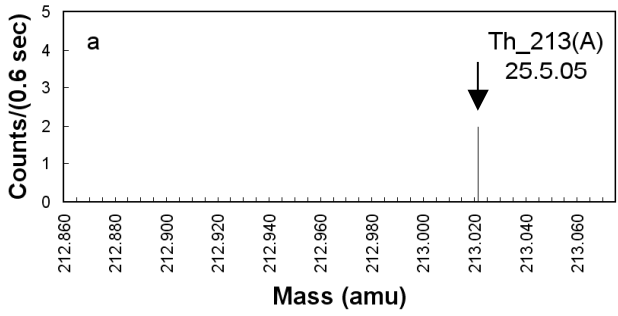

$M_{exp.} = 213.021$

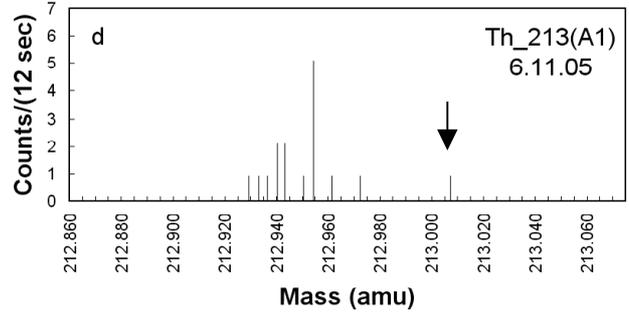

$M_{exp.}(c.m.) = 212.948$     $M_{exp.} = 213.007$
$M(^{197}Au^{16}O) = 212.961$    $M_{exp.}(corr.) = 213.020$

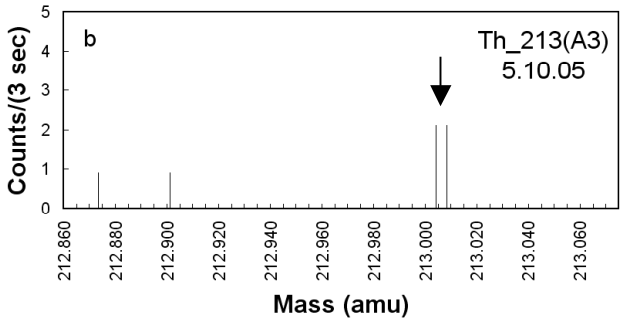

$M_{exp.}(c.m.) = 213.006$

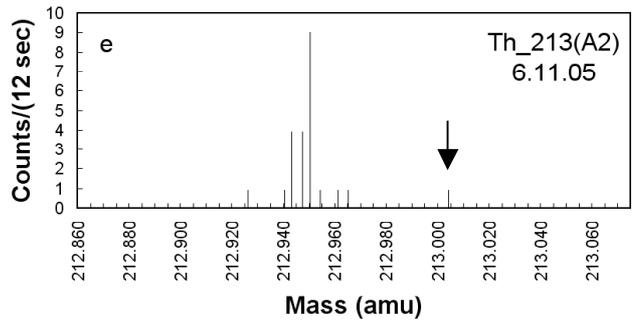

$M_{exp.}(c.m.) = 212.948$     $M_{exp.} = 213.004$
$M(^{197}Au^{16}O) = 212.961$   $M_{exp.}(corr.) = 213.017$

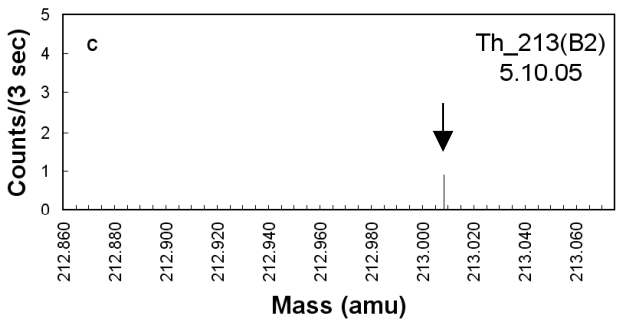

$M_{exp.} = 213.008$

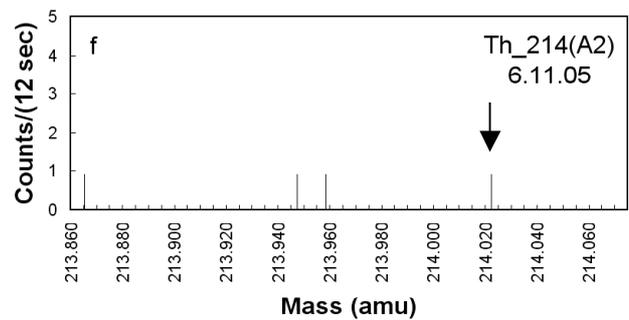

$M_{exp.}(c.m.) = 213.953$     $M_{exp.} = 214.022$
$M(^{198}Hg^{16}O) = 213.962$

$M(^{213}Th) = 213.013$            $M(^{214}Th) = 214.012$

Marinov_fig3

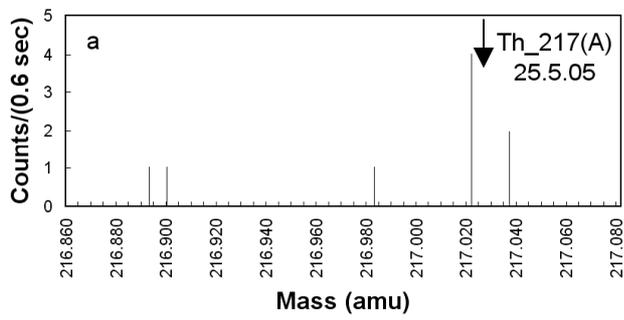

$M_{exp.}(c.m.) = 217.027$

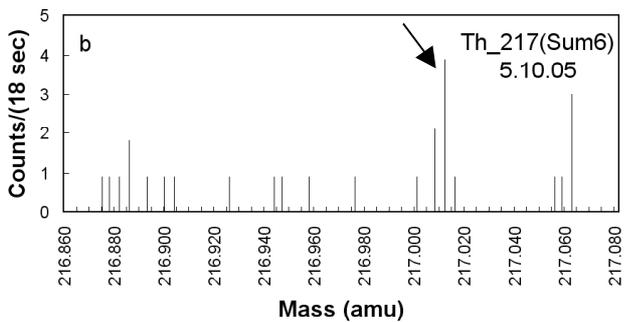

$M_{exp.}(c.m.) = 217.010$

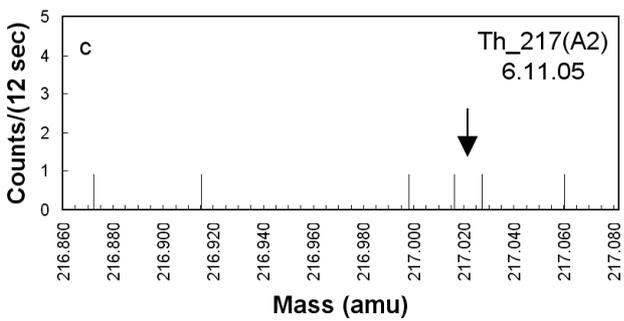

$M_{exp.}(c.m.) = 217.022$

$M(^{217}Th) = 217.013$

Marinov_fig4

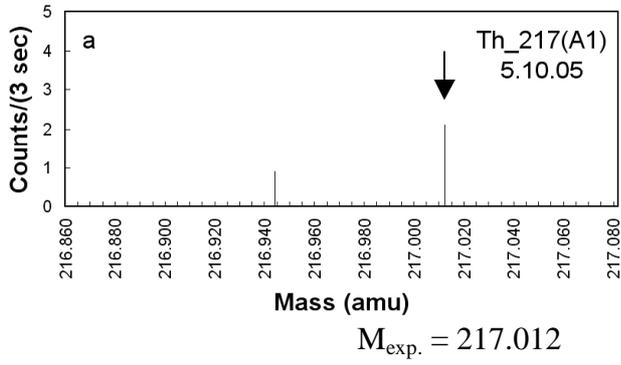
$M_{exp.} = 217.012$

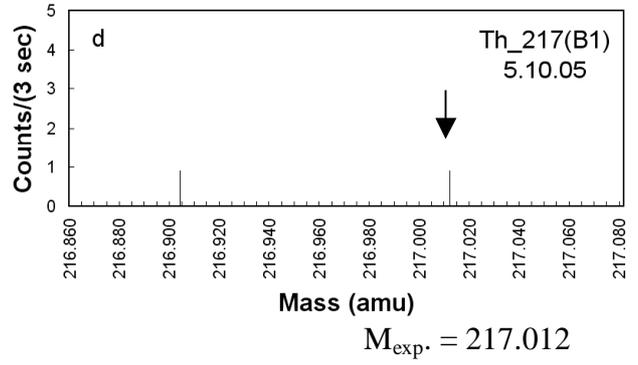
$M_{exp.} = 217.012$

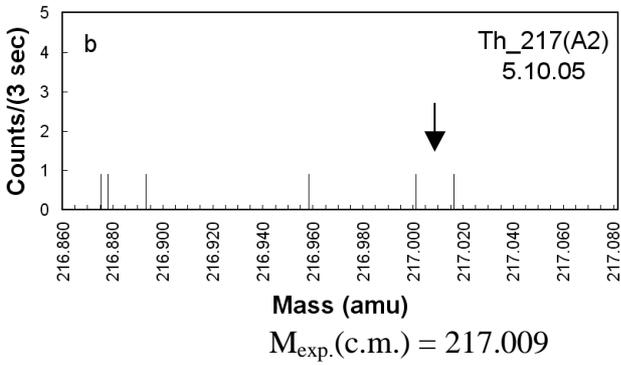
$M_{exp.}(c.m.) = 217.009$

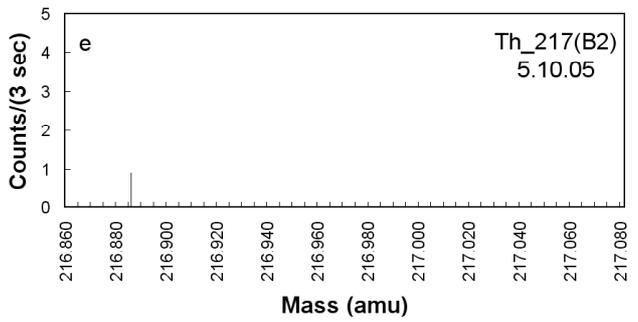

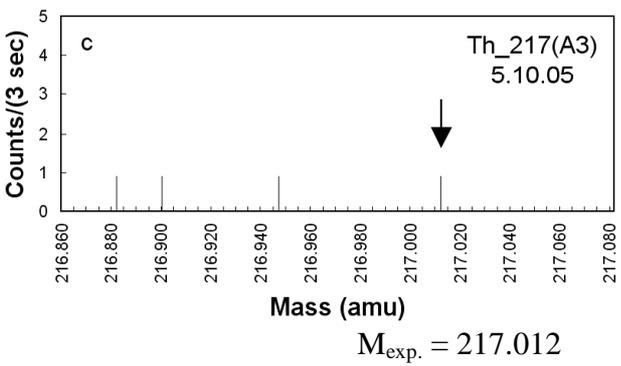
$M_{exp.} = 217.012$

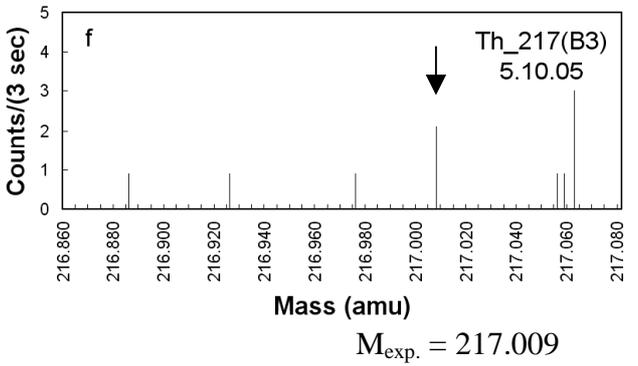
$M_{exp.} = 217.009$

$M(^{217}Th) = 217.013$

Marinov_fig5

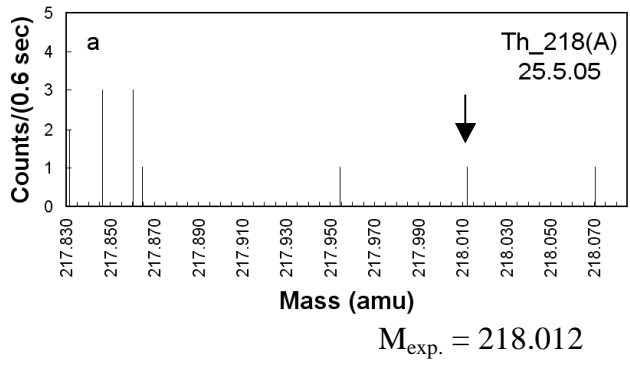

$M_{exp.} = 218.012$

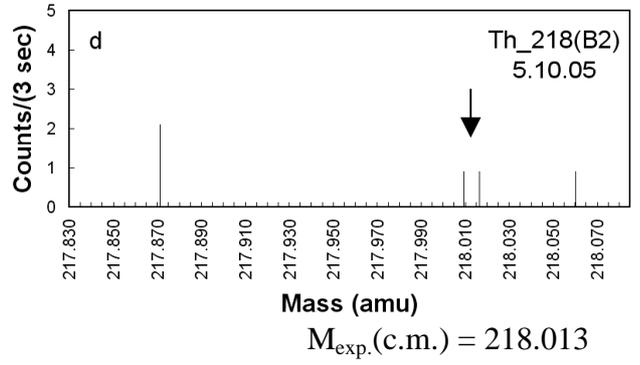

$M_{exp.}(c.m.) = 218.013$

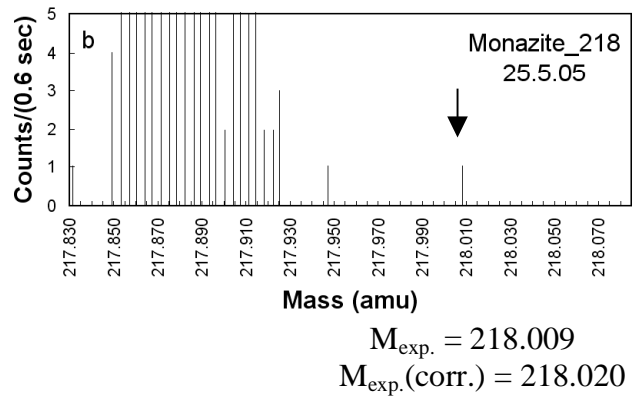

$M_{exp.} = 218.009$
$M_{exp.}(corr.) = 218.020$

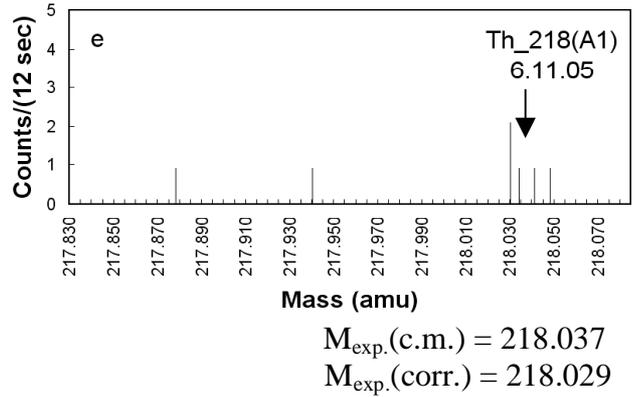

$M_{exp.}(c.m.) = 218.037$
$M_{exp.}(corr.) = 218.029$

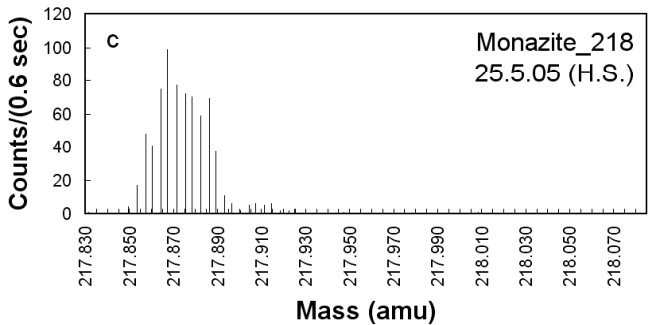

$M_{exp.}(c.m.) = 217.873$
$M(^{162}Dy^{40}Ar^{16}O) = 217.884$

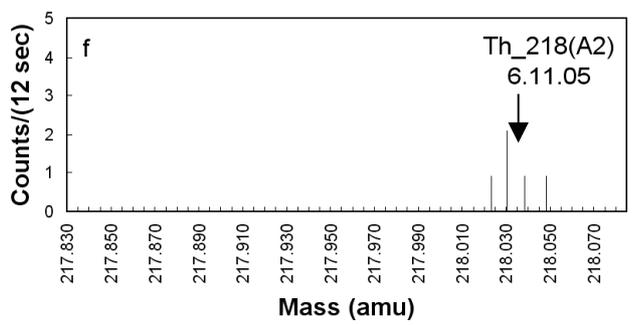

$M_{exp.}(c.m.) = 218.034$
$M_{exp.}(corr.) = 218.021$

$M(^{218}Th) = 218.013$

Marinov_fig6